\documentclass[submission, Proceedings]{SciPost}

\hypersetup{
    colorlinks,
    linkcolor={red!50!black},
    citecolor={blue!50!black},
    urlcolor={blue!80!black}
}

\usepackage[bitstream-charter]{mathdesign}
\DeclareSymbolFont{usualmathcal}{OMS}{cmsy}{m}{n}
\DeclareSymbolFontAlphabet{\mathcal}{usualmathcal}
\usepackage{physics}
\usepackage{dsfont}
\usepackage{booktabs}
\usepackage{subcaption}
\usepackage{tikz}
\usetikzlibrary{calc,decorations.markings,decorations.pathmorphing}
\usetikzlibrary{shapes.geometric}

\newcommand{\kpi}{\pi K}
\newcommand{\keta}{\eta K}
\newcommand{\ketap}{\eta^\prime K}

\newcommand{\TR}{T_\text{R}}

\newcommand{\disc}{\text{disc}\,}
\newcommand{\FF}{f_\text{s}}
\newcommand{\Ka}{K_0^*(700)}
\newcommand{\Kb}{K_0^*(1430)}
\newcommand{\Kc}{K_0^*(1950)}
\newcommand{\Kd}{K^*(892)}
\newcommand{\Ke}{K^*(1410)}

\newcommand{\mk}{M_K}
\newcommand{\meta}{M_\eta}

\newcommand{\mtau}{m_\tau}
\newcommand{\mr}[1]{\widetilde{M}_{#1}}

\newcommand{\beq}{\begin{equation}}
\newcommand{\eeq}{\end{equation}}

\newcommand{\Id}{\mathds{1}}
\newcommand{\GeV}{\,\text{GeV}}

\begin{document}

\begin{center}{\Large \textbf{
On the scalar $\boldsymbol{\pi K}$ form factor beyond the elastic region
}}\end{center}

\begin{center}
Frederic No\"el\textsuperscript{1$\star$},
Leon von Detten\textsuperscript{2},
Christoph Hanhart\textsuperscript{2}, 
Martin Hoferichter\textsuperscript{1},
Bastian Kubis\textsuperscript{3}
\end{center}

\begin{center}
{\bf 1} Albert Einstein Center for Fundamental Physics, Institute for Theoretical Physics, University of Bern, Sidlerstrasse 5, 3012 Bern, Switzerland
\\
{\bf 2} Forschungszentrum J\"ulich, Institute for Advanced Simulation, Institut f\"ur Kernphysik, and
J\"ulich Center for Hadron Physics, 52425 J\"ulich, Germany
\\
{\bf 3} Helmholtz-Institut f\"ur Strahlen- und Kernphysik and Bethe Center for Theoretical Physics, Universit\"at Bonn, 53115 Bonn, Germany
\\
$\star$ noel@itp.unibe.ch
\end{center}

\begin{center}
November 16, 2021
\end{center}


\definecolor{palegray}{gray}{0.95}
\begin{center}
\colorbox{palegray}{
  \begin{minipage}{0.95\textwidth}
    \begin{center}
    {\it  16th International Workshop on Tau Lepton Physics (TAU2021),}\\
    {\it September 27 – October 1, 2021} \\
    \doi{10.21468/SciPostPhysProc.?}\\
    \end{center}
  \end{minipage}
}
\end{center}

\section*{Abstract}
{\bf 
Pion--kaon ($\boldsymbol{\pi K}$) final states, often appearing in heavy-particle decays at the precision frontier, are important for Standard-Model tests, to describe crossed channels with exotic states, and for spectroscopy of excited kaon resonances. 
We construct a representation of the $\boldsymbol{\pi K}$ $\boldsymbol{S}$-wave form factor using the elastic $\boldsymbol{\pi K}$ scattering phase shift via dispersion relations in the elastic region and extend this model into the inelastic region using resonance exchange, while maintaining unitarity and the correct analytic structure.  
As a first application, we successfully described the $\boldsymbol{\tau \to K_S \pi \nu_\tau}$ spectrum to not only achieve a better distinction between $\boldsymbol{S}$- and $\boldsymbol{P}$-wave contributions, but also to provide an improved estimate of the $\boldsymbol{CP}$ asymmetry produced by a tensor operator as well as the forward--backward asymmetry, both of which can be confronted with future data at Belle  II. The work presented here is published in Ref.~\cite{vonDetten:2021euh}.
}

\section{Introduction}
\label{sec:intro}
When searching for physics beyond the Standard Model (BSM), including in $CP$-violating observables, one often encounters multi-hadron final states, e.g., in semi-leptonic, $D$- and $B$-meson decays. In addition, the identification of exotic resonances in such final states requires control over rescattering effects.
Especially for heavy particles decaying with net strangeness, it thus becomes increasingly important to also describe the abundantly appearing final-state interactions of kaons and pions up to high energies. In particular a consistent description of $\pi K$ scattering and production can serve as a test of SM physics, be used to search for exotic hadronic states in crossed channels, and improve the spectroscopy of excited kaon resonances. 

To be more specific, in the quest for  $CP$ violation beyond the SM the inelastic contributions to the $\pi K$ channel enter in the $CP$ asymmetry in $\tau \to K_S \pi \nu_\tau$ generated by a tensor operator, as the elastic contributions cancel  by Watson's theorem~\cite{Cirigliano:2017tqn}. Further, in the hunt for exotic hadrons the $Z_c(4430)$ was discovered by Belle and LHCb in the reaction $B \to \psi^\prime \pi K$ in the $\psi^\prime \pi$ subsystem~\cite{Mizuk:2009da,Aaij:2014jqa}. Since, to describe such a crossed process, the different partial waves of the $\pi K$ subsystem interfere, a high control over especially their phases is compulsory, which cannot be achieved by a simple Breit--Wigner (BW) model. 
Hence, generally speaking a better understanding of the $\pi K$ form factors is needed to describe all these processes appropriately. 

For these purposes, we constructed a representation of the $\pi K$ $S$-wave form factor using the elastic $\pi K$ scattering phase shifts via dispersion relations in the elastic region, as demanded by Watson's theorem, and extended this model into the inelastic region using resonance exchange, while maintaining unitarity and the correct analytic structure~\cite{Hanhart:2012wi,Ropertz:2018stk}.
As a first application, we successfully described the ${\tau \to K_S \pi \nu_\tau}$ spectrum, including the highly overlapping $S$-wave resonance $K^*_0(1430)$ and $P$-wave resonance $K^*(1410)$. In contrast to common BW parameterizations, which violate unitarity, our parameterization has the correct phase behavior built in and fulfills unitarity by construction. 
For an improved separation of these resonances using future measurements, we further calculated forward--backward (FB) asymmetries for the different fit scenarios. In addition, we could use our results to refine the estimate of the $CP$ asymmetry generated by a tensor operator. Finally, we were able to extract the resonance properties of the $K^*_0(1430)$ and $K^*_0(1950)$ via Pad\'e approximants. 
 Here, we provide a summary of the main ideas and applications, while deferring 
a more detailed discourse to Ref.~\cite{vonDetten:2021euh}. 

\section{Formalism} \label{sec:form}

Our formalism for the $T$-matrix that fulfills the criteria of unitarity and analyticity is built upon the Bethe--Salpeter equation, which in channel space in matrix form reads 
\begin{equation}
T_{if}=V_{if}+V_{im} G_{mm} T_{mf}\,,
\label{eq:T_bethe_salpeter}
\end{equation}
and fulfills unitarity as long as $V_{if}\in\mathbb{R}$ and $\disc G_{mm} = 2 i \rho_m$, where $\rho_m$ denotes the two-body phase space in channel $m$. Furthermore, we employ the so-called two-potential formalism~\cite{Nakano:1982bc}, which starts by splitting the scattering  potential $V$ into two pieces,
\begin{equation}
V= V_0 + V_\text{R}\,.
\label{eq:V_split}
\end{equation}
Accordingly, this results in a corresponding splitting of the $T$-matrix
\begin{equation}
T=T_0+\TR\,,
\label{eq:T_split}
\end{equation}
where $T_0$ fulfills the Bethe--Salpeter equation
that has $V_0$ as input, ${T_0=V_0+V_0 G T_0}$.
In our application we assume $T_0$ to be purely elastic and consider the additional channels to couple only through the resonance exchange in $T_\text{R}$, as motivated by the phenomenologically successful isobar model~\cite{Anisovich:2002ij,Klempt:2007cp,Battaglieri:2014gca}. 
Employing a two-channel setup, corresponding in our application to $\kpi$ and $\ketap$, we therefore have
\begin{equation}
T_0=\begin{pmatrix}
\tfrac{1}{\rho_1}\sin{\delta_0} e^{i\delta_0} & 0\\
0 & 0
\end{pmatrix}\,,
\end{equation}
which only depends on the scattering phase $\delta_0$ and makes any explicit parameterization of $V_0$ obsolete, using an empirical parameterizations of $\delta_0$ instead.

By means of dispersion theory we can use the given constraints to calculate $T_\text{R}$ and consequently the full scattering $T$-matrix, which is given as
\begin{equation}
T = T_0 + T_\text{R} = T_0 + \Omega \left[ \Id - V_\text{R} \Sigma \right]^{-1} V_\text{R} \Omega^\text{T}\,,
\label{eq:T_theory}
\end{equation}
with 
\begin{equation}
\label{eq:Omnes}
\Omega=\begin{pmatrix}
\Omega_{11} & 0\\
0 & 1
\end{pmatrix}, \qq{} \Omega_{11}=\exp \qty(\frac{s}{\pi} \int_{s_\text{th}}^\infty \text{d}z \frac{\delta_0(z)}{z(z-s)})\,,
\end{equation}
the Omn\`es function for the $\kpi$ channel and
\begin{equation}
\Sigma_{ij}(s)=\frac{s}{2 \pi i } \int_{s_\text{th}}^{\infty} \text{d}z \frac{\Omega^\dagger_{im}(z)  \disc G_{mm}(z) \Omega_{mf}(z)}{z(z-s)}\,,
\label{eq:selfenergy}
\end{equation}
the dressed loop operator also called self energy. Furthermore, we parameterize the resonance potential $V_\text{R}$ as
\begin{align}
V_\text{R}(s)_{ij}
&= \sum_r g_i^{(r)}\frac{s-s_0}{\qty(s-\mr{(r)}^2)\qty(s_0-\mr{(r)}^2)} g_j^{(r)}\,, \label{eq:VR}
\end{align}
which is subtracted at ${s_0=(\mk+\meta)^2}$ to remove low-energy contributions already considered via $T_0$. Here the mass $\mr{(r)}$ is the bare mass of resonance $r$ and $g_i^{(r)}$ is the bare coupling of the resonance $r$ to channel $i$. 
The scalar form factor $\FF$, given in channel space via
\begin{align}
	\qty(\FF)_i= M_i + T_{im} G_{mm} M_m\,, \label{eq:FF0}
\end{align}
can now be expressed as
\begin{equation}
\FF(s) = \Omega(s) \left[  \Id-V_\text{R}(s) \Sigma(s)  \right]^{-1} M(s) \,,
\label{eq:FF}
\end{equation} 
where $M$ is a reparameterized source term, which can be written as
\begin{align}
M_i &= \sum_{k=0}^{k_{\rm max}} c_i^{(k)}s^k - \sum_{r} g_i^{(r)} \frac{s-s_0}{\qty(s-\mr{(r)}^2)\qty(s_0-\mr{(r)}^2)} \alpha^{(r)} \,.
\label{eq:FF_source}
\end{align}
Here, the coefficients $c_i^{(k)}$ and the resonance couplings $\alpha^{(r)}$ are parameters of the model that depend on the source.

\section{Fit to scattering data} \label{sec:scat}

We fit our parameterization of the $\pi K$ ${I=\frac{1}{2}}$ $S$-wave in combination with an elastic ${I=\frac{3}{2}}$ parameterization from Ref.~\cite{Pelaez:2016tgi} to the data of Ref.~\cite{Aston:1987ir}. 
\begin{figure}[t]%
	\includegraphics[width=0.65\linewidth,trim = 30pt 30pt 0pt 0pt]{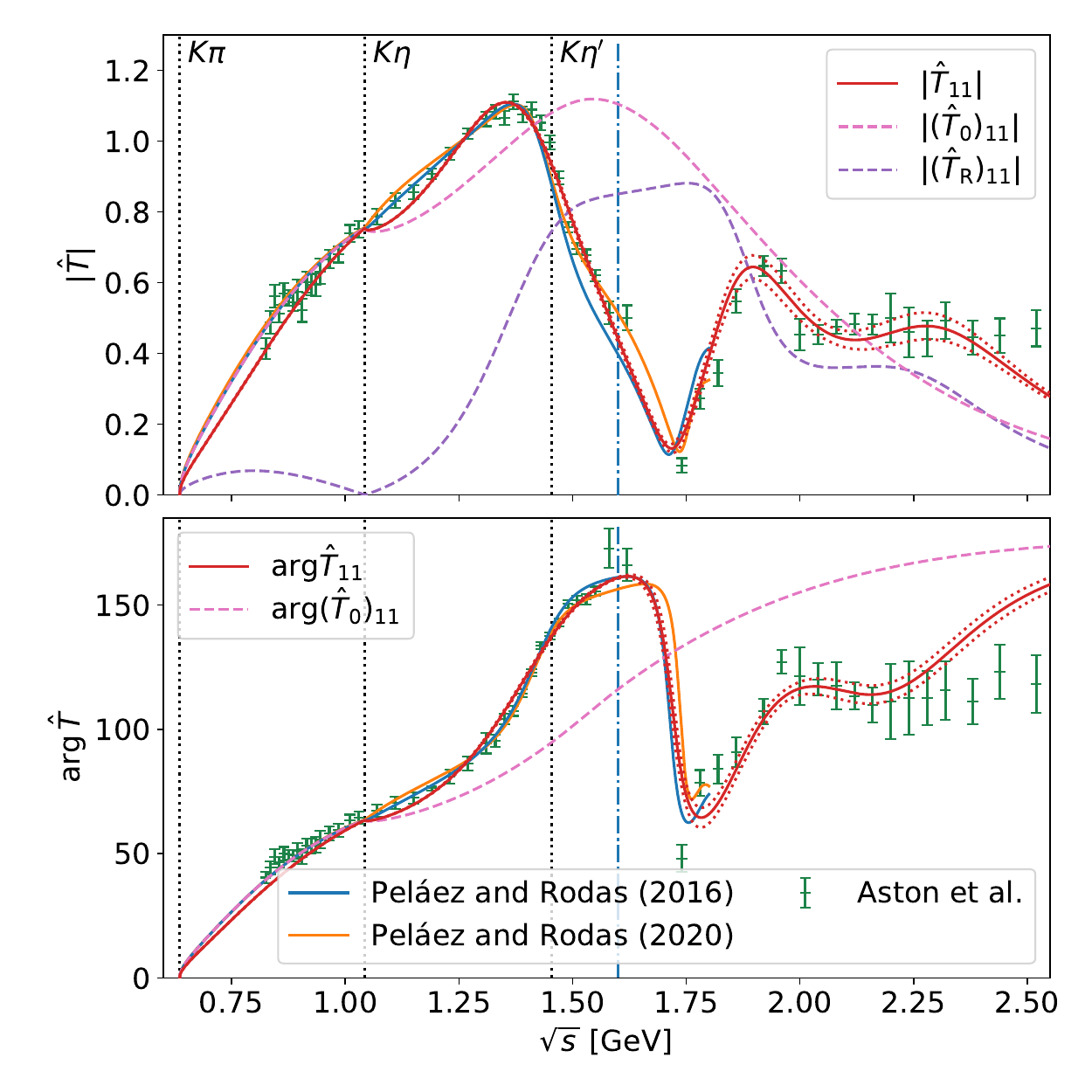}
	\centering
	\caption{
		Plot of the combined fit of argument and absolute value of ${\hat{T}_{if}=\rho_i  \big(T^{\frac{1}{2}} + T^{\frac{3}{2}}/2\big)_{if}}$, with 1$\sigma$ uncertainty band, to the corresponding data of Aston et al.~\cite{Aston:1987ir} in comparison to the results of Pel\'aez and Rodas~\cite{Pelaez:2016tgi,Pelaez:2020gnd}. We further show the accordingly defined low-energy amplitude $\hat{T}_0$ and the resonance part of the model $\hat{T}_\text{R}$, independently.
	}
	\label{fig:Fit_Aston}
\end{figure}%
As phase input we use the elastic part of Ref.~\cite{Pelaez:2016tgi}, where we turn off all resonant contributions and guide the phase smoothly to $\pi$ at ${\sqrt{s_\text{m}}=1.52\GeV}$, thus only including the lowest resonance $\Ka$ within the phase. As we aim at a description from the $\kpi$ threshold up to $2.5\GeV$ the additional scalar resonances $\Kb$ and $\Kc$ are included explicitly via $V_\text{R}$. We consider a two-channel setup, incorporating only the $\kpi$ and $\ketap$ channel, as the $\keta$ channel  turns out empirically to effectively decouple in that energy range. Figure \ref{fig:Fit_Aston} shows the results of the combined fit of argument and absolute value. Considering the modest quality of the data, we find the fit suitable up to about $2.3\GeV$. An extension to higher energies would require the inclusion of yet another $K_0^*$ resonance, which however would require reliable data up to even higher energies. 

\section{Application to $\boldsymbol{\tau}$ decays} \label{sec:tau}

Using the scattering parameterization acquired in the previous section we now have, using Eq.~\eqref{eq:FF}, a parameterization of the $\kpi$ ${I=\frac{1}{2}}$ scalar form factor, with free parameters only contained within the source term~\eqref{eq:FF_source}. For now we use this parameterization to describe the scalar form factor within the decay spectrum of $\tau^- \to K_S \pi^- \nu_\tau$ as measured by the Belle collaboration \cite{Epifanov:2007rf}. 
We fix the scalar form factor using our parameterization via $(f_s)_1$, see Eq.~\eqref{eq:FF}, while we use for the vector form factor a more conventional parameterization from resonance chiral perturbation theory~\cite{Moussallam:2007qc,Boito:2008fq,Boito:2010me,Bernard:2011ae,Antonelli:2013usa}, where we follow the conventions from Ref.~\cite{Antonelli:2013usa}. There we choose the subtraction constants fixed to their central values (as determined independently from $K_{l3}$ decays) and allow the bare mass and width as well as the mixing parameter to vary in such a way that the shape of the generated $\kpi$ $P$-wave remains phenomenologically viable.

\begin{figure}[t]%
	\includegraphics[width=0.75\linewidth,trim = 30pt 30pt 0pt 0pt]{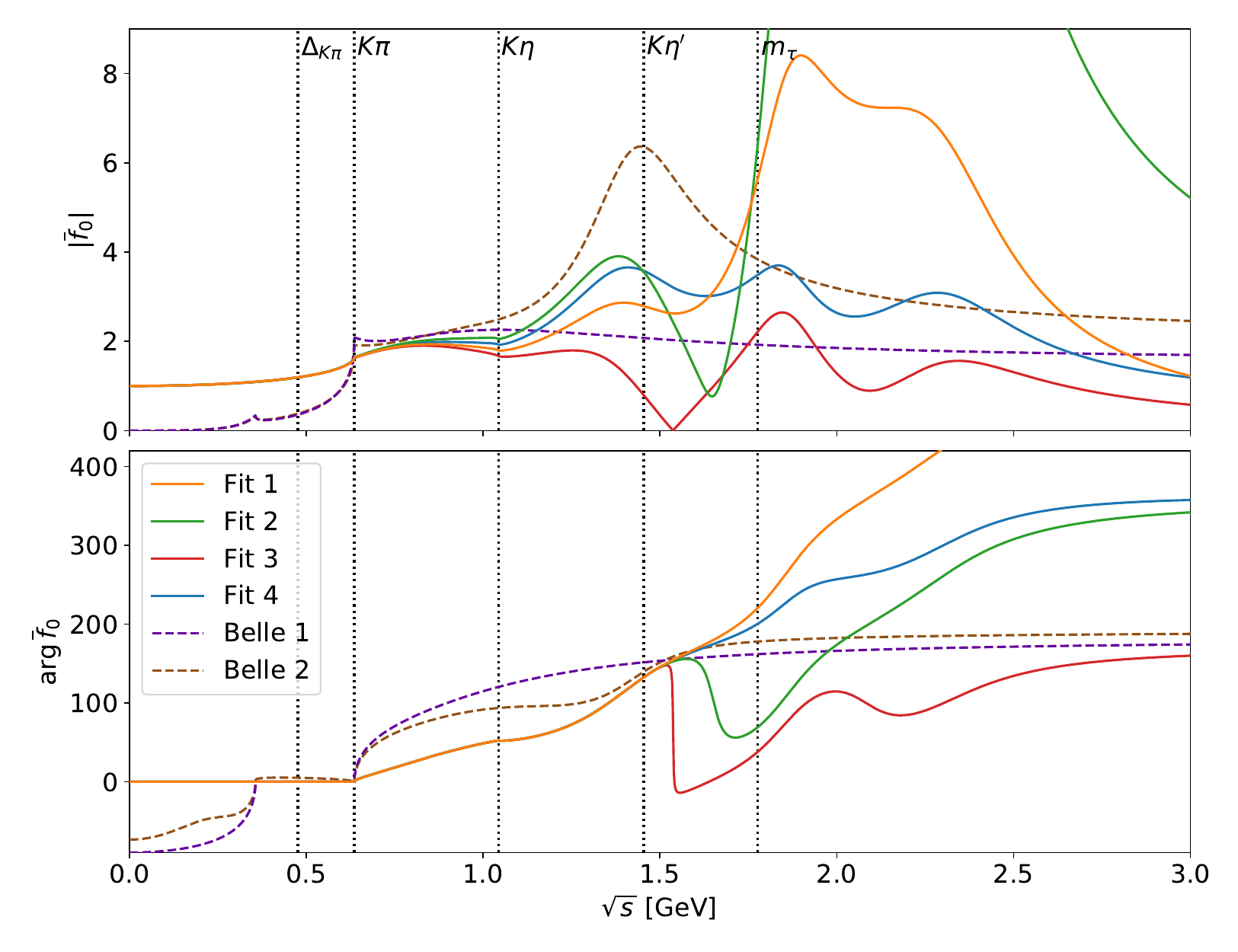}
	\centering
	\caption{Scalar form factor $\bar{f}_0$ for the fit of our parameterization of the total decay rate of ${\tau^- \rightarrow K_S \pi^- \nu_\tau}$ to efficiency-corrected and background-reduced events of Ref.~\cite{Epifanov:2007rf} in comparison to two BW parameterizations ``Belle 1'' and ``Belle 2''~\cite{Epifanov:2007rf}---which include $\Ka$, $\Kd$, and $\Ke$, or $\Ka$, $\Kd$, and $\Kb$, respectively. We consider different combinations of fixing the $\alpha^{(2)}$ and $c^{(1)}_1$ parameters. $c^{(0)}_1$ and $c^{(0)}_2$ are implicitly determined by the normalization via $(f_s)_1(0)=1$ and by $U(3)$ ChPT via $(f_s)_2(0)=\sqrt{3}$, respectively. The Callan--Treiman low-energy theorem is also included as an additional constraint into the fit.
	}
	\label{fig:fit_SFF}
\end{figure}%
We find four representative fit scenarios, which turn out to be visibly indistinguishable in the total decay rate. However, the actual scalar form factors contained within the four parameterizations, see Fig.~\ref{fig:fit_SFF}, differ more substantially, especially above the $\eta'K$ threshold, where the contribution to the decay rate is very suppressed by the phase space and the data points have large uncertainties. All fits are equally well suitable to describe the decay rate and have similar fit statistics with a reduced $\chi^2 \approx 1$. 
Despite the strong overlap of the $K^*(1410)$ and $K^*_0(1430)$ in the total decay rate, which in the past was usually solved by discarding one of the resonances, we were able to get distinct parameters for both resonances, as in our application the resonance couplings to the $K^*_0(1430)$ are already fixed by the scattering data.
The $K^*_0(1950)$, on the other hand, is difficult to constrain from this fit, as it lies above the $\tau$ mass and its influence on the decay region is quite limited, which can also be seen by the huge contributions in the scalar form factor in Fit~1 and 2, above the boundary of the phase space with a free $\alpha^{(2)}$. 
However, even the parameterizations without a source term coupling to the $K^*_0(1950)$ (Fit~3 and 4 with $\alpha^{(2)}=0$) show resonant structures around its mass, which nicely shows the built-in unitarity indirectly including the knowledge of the scattering phase about all resonances. 
Furthermore, by construction the phase of the scalar form factor is fixed up to the $\eta'K$ threshold to the scattering phase as demanded by Watson's theorem and elastic unitarity. 
This is in marked contrast to the BW parameterizations by Belle~\cite{Epifanov:2007rf}, which show very inconsistent phases at low energies and unphysical structures below the $\pi K$ threshold. 
Including a linear term $c^{(1)}_1$ in the source term (Fit~2 and 4) gives some slight improvements to the fit quality, but changes the high-energy behavior of the scalar form factor to approach a constant instead of falling of like $1/s$, as expected by perturbative-QCD arguments. As the fit statistics are even without this linear term sufficient, we conclude that all fits are basically equivalent. To further distinguish the different fit scenarios, additional data are required.

\section{Implications and results}

\subsection{Forward--backward asymmetry}
\begin{figure}[tp]%
	\includegraphics[width=0.5\linewidth,trim = 30pt 30pt 0pt 0pt]{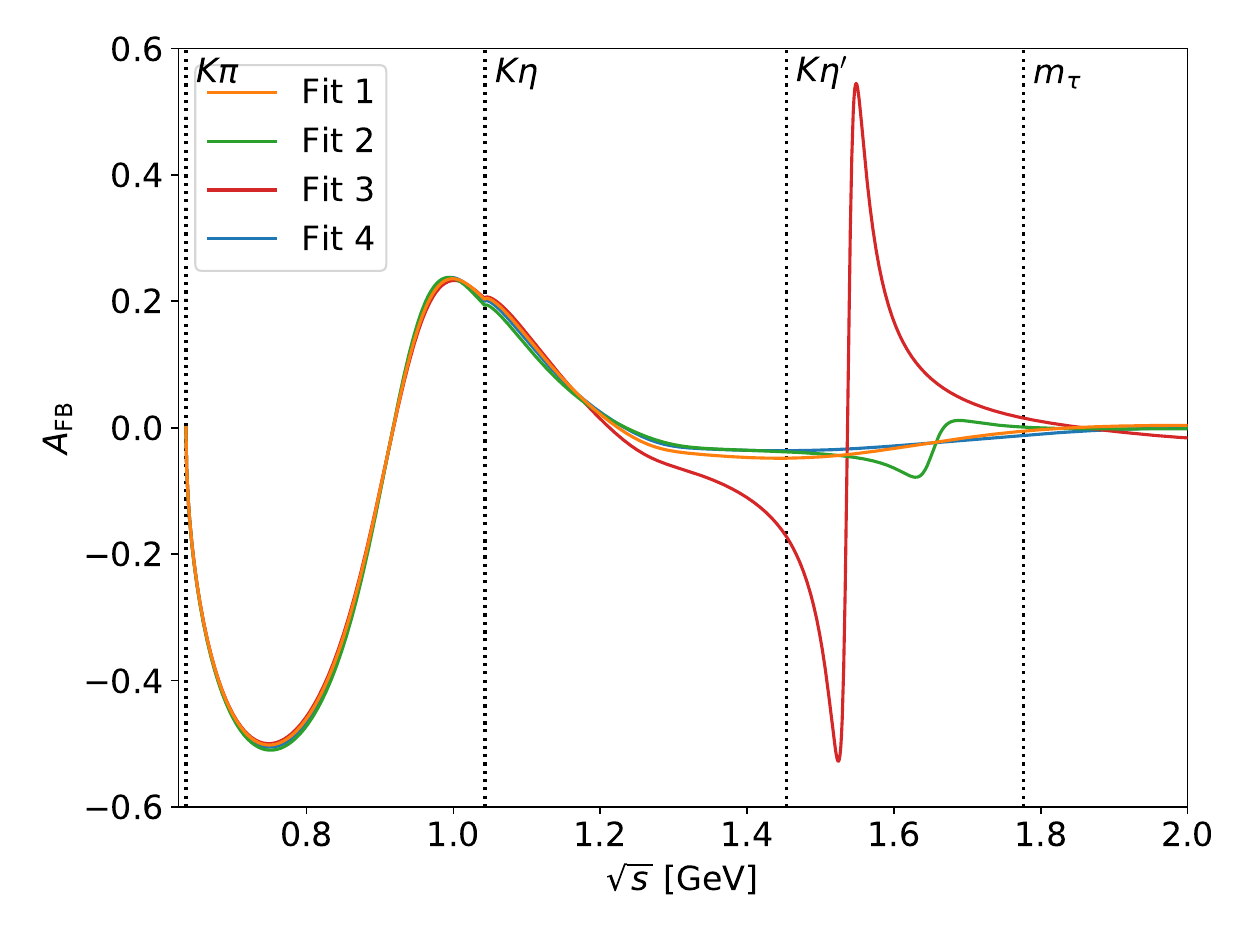}
	\centering
	\caption{FB asymmetry as defined in Eq.~\eqref{eq:A_FB} for the four fit variants.}
	\label{fig:AsymFB}
\end{figure}%
One such possibility to further distinguish the different fit scenarios would be to examine the FB asymmetry~\cite{Beldjoudi:1994hi,Kou:2018nap} as can be measured by Belle II~\cite{Kou:2018nap}:
\begin{align}
	A_\text{FB}(s)&= \frac{\int_0^1 \text{d}z\left[\frac{\text{d}\Gamma}{\text{d}z}(z)-\frac{\text{d}\Gamma}{\text{d}z}(-z)\right]}{\int_0^1 \text{d}z\left[\frac{\text{d}\Gamma}{\text{d}z}(z)+\frac{\text{d}\Gamma}{\text{d}z}(-z)\right]} 
	=\frac{- 2 \Re(f_0 f_+^*) \Delta_{\kpi} q_{\kpi} \sqrt{s}}{|f_0|^2 \Delta_{\kpi}^2 + \frac{4}{3} |f_+|^2 q_{\kpi}^2 \qty(\frac{2 s^2}{\mtau^2} + s)}\,,\label{eq:A_FB}
\end{align}
where $z$ denotes the cosine of the $\pi K$ helicity angle. This quantity would separate vector and scalar components and make it easier to distinguish the $K^*(1410)$ and $K^*_0(1430)$ contributions. We show the FB asymmetry for the four fit scenarios in Fig.~\ref{fig:AsymFB}. Major differences can be seen above the $\eta'K$ threshold, as expected due to the different phase motion in that energy region. 

\subsection{Branching ratio}
By integrating over the differential decay rate we calculate the branching ratio for $\tau\to K_S \pi \nu_\tau$. Averaging over all four fit scenarios with its spread as systematic uncertainty (sys) we get
\begin{align}
	\text{BR}({\tau\rightarrow K_S \pi \nu_\tau}) &=
	4.35(6)_\text{st}(3)_\text{norm}(7)_\text{sys}\times10^{-3}
	=4.35(10)\times10^{-3}\,,
\end{align}
where we also included the statistical error (st) propagated from the fit parameters and the uncertainty from the normalization constants (norm). Although this result is $2\sigma$ above the original Belle result $\text{BR}({\tau\rightarrow K_S \pi \nu_\tau})|_\text{\cite{{Epifanov:2007rf}}}=4.04(13)\times 10^{-3}$, it agrees at $1.5\sigma$ with the more recent ${\text{BR}({\tau\rightarrow K_S \pi \nu_\tau})|_\text{\cite{{Ryu:2014vpc}}} = 4.16(8)\times 10^{-3}}$ as well as the Particle Data Group average \linebreak${\text{BR}({\tau\rightarrow K_S \pi \nu_\tau})|_\text{\cite{{Zyla:2020zbs}}} = 4.19(7)\times 10^{-3}}$.

\subsection{$\boldsymbol{CP}$ asymmetry}
We were further able to improve the estimate of the BSM $CP$ asymmetry produced by a tensor operator with Wilson coefficient $c_T$ interfering with the vector operator. Due to the absence of a scalar--vector interference this is the only option to generate a $CP$ asymmetry with new heavy degrees of freedom. As shown in Ref.~\cite{Cirigliano:2017tqn}, by Watson's theorem it follows that in this case all elastic contributions to the $CP$ asymmetry cancel identically, as vector and tensor operators follow the same unitarity condition and thus have to have the same strong phase in the purely elastic case. Hence inelastic effects,  included in our parameterization via resonance exchange, are mandatory to obtain a non-vanishing $CP$ asymmetry. Within the SM, a $CP$ asymmetry is generated by $K^0$--$\bar K^0$ mixing, but the corresponding  prediction shows a $2.8\sigma$ tension with the 2012 BaBar measurement~\cite{BABAR:2011aa}, which could point to $CP$ violation beyond the SM. 
Using our parameterization we find 
\begin{align}
	\label{CPasym}
	A_{CP}^{\tau,\text{BSM}} = -0.034(14)\,\Im c_T\,,
\end{align}
which supports the simpler estimate of Ref.~\cite{Cirigliano:2017tqn}. Unfortunately, limits on $\Im c_T$ from the neutron
electric dipole moment and $D$--$\bar{D}$ mixing rule out this mechanism to explain the tension.

\subsection{Pole extraction}
\begin{figure}[tp]%
	\begin{subfigure}{0.475\textwidth}
		\includegraphics[width=\linewidth,trim = 18pt 19pt 0pt 0pt]{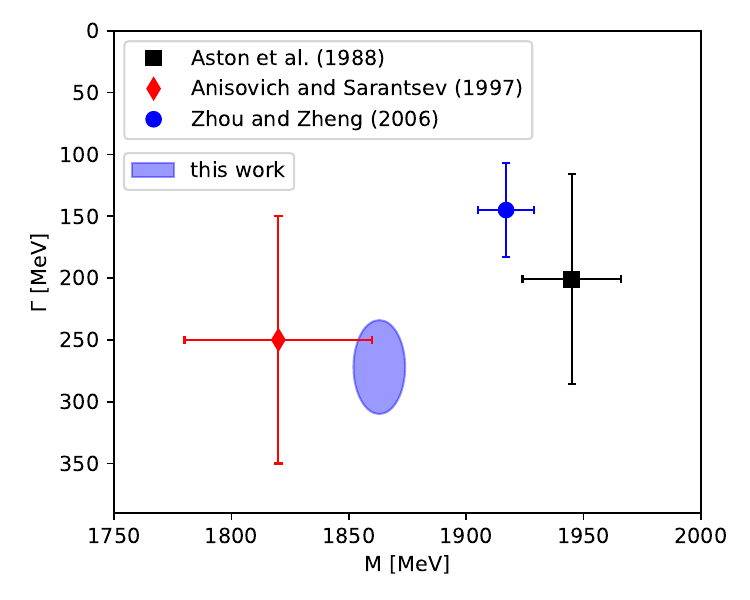}
		\centering
	\end{subfigure}
	\hfill
	\begin{subfigure}{0.475\textwidth}
		\includegraphics[width=\linewidth,trim = 18pt 19pt 0pt 0pt]{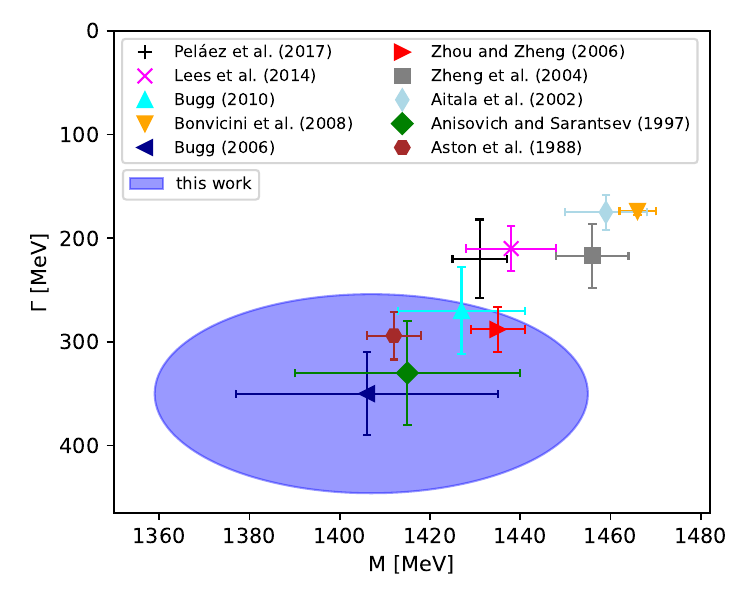}
		\centering
	\end{subfigure}
	\caption{Extracted pole position of the $\Kc$ (left) and $\Kb$ (right) in comparison to the works of Aston et al.~\cite{Aston:1987ir}, Anisovich and Sarantsev~\cite{Anisovich:1997qp}, and Zhou and Zheng~\cite{Zhou:2006wm} as well as Pel\'aez et al.~\cite{Pelaez:2016klv}, Lees et al.~\cite{Lees:2014iua}, Bugg~(2010)~\cite{Bugg:2009uk}, Bonvicini et al.~\cite{Bonvicini:2008jw}, Bugg~(2006)~\cite{Bugg:2005xx}, Zhou  and Zheng~\cite{Zhou:2006wm}, Zeng et al.~\cite{Zheng:2003rw}, Aitala et al.~\cite{Aitala:2002kr}, Anisovich and Serantsev~\cite{Anisovich:1997qp}, and Aston et al.~\cite{Aston:1987ir}, respectively.}
	\label{fig:poles}
\end{figure}%
We were able to extract the pole positions as well as residues of the two scalar resonances $\Kb$ and $\Kc$ using Padé approximants as shown in Fig.~\ref{fig:poles}. For both resonances our results are in reasonable agreement with previous extractions, indicating a $\Kb$ mass towards the lower end, see Ref.~\cite{vonDetten:2021euh} for more details on the extraction of the pole parameters and residues.  

Furthermore, we extracted the coupling of the $R=K^*_0(1430)$ resonance to the $\bar{s}\gamma^\mu u$ current in a model-independent way via its residue $C_\text{R}^{us}$, which can then be re-interpreted in a narrow-width sense as a decay rate 
\begin{align}
\label{tauK0BR}
    \Gamma(\tau \to \text{R} \nu_\tau) 
    & = \frac{ 6 \pi^2 c_\Gamma \Delta_{\kpi}^2}{M_\text{R}^4} \bigg(1-\frac{M_\text{R}^2}{m_\tau^2}\bigg)^2\big|C_\text{R}^{us}\big|^2.
\end{align}
We find an upper bound on the branching ratio of $\text{BR}(\tau\to \Kb \nu_\tau) < 1.6 \times 10^{-4}$ (at $95\%$ confidence level), which is by a factor of 3 better than the current literature values~\cite{Zyla:2020zbs,Barate:1999hj}. 
Our residues (from the four fit scenarios) are scattered around the values from other theoretical investigations, including Refs.~\cite{Maltman:1999jn,ElBennich:2009da}, which however are more rigid in the fit function. For instance, Ref.~\cite{ElBennich:2009da} uses a coupled-channel Omn\`es matrix, which requires inputs for scattering phases and elasticity parameters for all channels that are not available at the moment. The implementation from Ref.~\cite{ElBennich:2009da} circumvents this issue by relying on further chiral constraints, with the scalar form factor uniquely determined from the $T$-matrix. Our parameterization, on the other hand, also fulfills the chiral constraints, but allows for more flexibility by adding further terms in the resonance potential. 

\section{Conclusion} \label{sec:con}
In conclusion, the formalism proposed in Ref.~\cite{vonDetten:2021euh} has proven adequate for the description of $\pi K$ $S$-wave scattering as well as the scalar form factor up to the $K^*_0(1950)$. It should thus also allow for a meaningful description of the $\pi K$ form factors in future analyses of semi-leptonic $D$- and $B$-meson decays and transfer to higher partial waves.
 Future measurements expected from Belle II, for the $\tau \to K_S \pi \nu_\tau$ spectrum, FB and $CP$ asymmetry, will improve the phenomenology presented here, especially in the inelastic region, and thereby provide valuable input for controlling $\pi K$ final-state interactions in more complicated systems. 

\paragraph{Funding information}
Financial support by the SNSF (Project No.\ PCEFP2\_181117),
the DFG through the funds provided to the Sino--German Collaborative
Research Center TRR110 ``Symmetries and the Emergence of Structure in QCD'' (DFG Project-ID 196253076 -- TRR 110),
the Bonn--Cologne Graduate School of Physics and Astronomy (BCGS),
and the European Union's Horizon 2020 research and innovation programme under grant agreement No.\ 824093
is gratefully acknowledged.

\bibliography{refs.bib}

\nolinenumbers

\end{document}